\documentclass[twocolumn]{aastex631}

\usepackage{chemformula}
\usepackage[version=4]{mhchem}

\turnoffediting

\begin{document}

\title{On the origin of infrared bands attributed to tryptophan in Spitzer observations of IC 348}

\author{Aditya Dhariwal}
\affiliation{Department of Chemistry, University of British Columbia, 2036 Main Mall, Vancouver BC V6T 1Z1, Canada}

\author{Thomas H. Speak}
\affiliation{Department of Chemistry, University of British Columbia, 2036 Main Mall, Vancouver BC V6T 1Z1, Canada}

\author{Linshan Zeng}
\affiliation{Department of Chemistry, University of British Columbia, 2036 Main Mall, Vancouver BC V6T 1Z1, Canada}
\author{Amirhossein Rashidi}
\affiliation{Department of Chemistry, University of British Columbia, 2036 Main Mall, Vancouver BC V6T 1Z1, Canada}
\author{Brendan Moore}
\affiliation{Department of Chemistry, University of British Columbia, 2036 Main Mall, Vancouver BC V6T 1Z1, Canada}

\author{Olivier Bern\'{e}}
\affiliation{Institut de Recherche en Astrophysique et Planétologie, Université de Toulouse, CNRS, CNES, UPS, Toulouse, France}

\author[0000-0001-9479-9287]{Anthony J. Remijan}
\affiliation{National Radio Astronomy Observatory, Charlottesville, VA 22903, USA}

\author{Ilane Schroetter}
\affiliation{Institut de Recherche en Astrophysique et Planétologie, Université de Toulouse, CNRS, CNES, UPS, Toulouse, France}

\author[0000-0003-1254-4817]{Brett A. McGuire}
\affiliation{Department of Chemistry, Massachusetts Institute of Technology, Cambridge, MA 02139, USA}
\affiliation{National Radio Astronomy Observatory, Charlottesville, VA 22903, USA}
\author[0000-0002-2887-5859]{V\'ictor M. Rivilla}
\affiliation{Centro de Astrobiolog{\i}a (CAB), INTA-CSIC, Carretera de Ajalvir km 4, Torrej{\'o}n de Ardoz, 28850 Madrid, Spain}
\author{Arnaud Belloche}
\affiliation{Max-Planck-Institut für Radioastronomie (MPIfR), Auf dem Hügel 69, 53121 Bonn, Germany.}

\author[0000-0001-9133-8047]{Jes K. J{\o}rgensen}
\affiliation{Niels Bohr Institute, University of Copenhagen, {\O}ster Voldgade 5-7, DK-1350 Copenhagen K., Denmark}

\author{Pavle Djuricanin}
\affiliation{Department of Chemistry, University of British Columbia, 2036 Main Mall, Vancouver BC V6T 1Z1, Canada}

\author[0000-0001-8976-1938]{Takamasa Momose}
\affiliation{Department of Chemistry, University of British Columbia, 2036 Main Mall, Vancouver BC V6T 1Z1, Canada}
\author[0000-0002-0850-7426]{Ilsa R. Cooke}
\affiliation{Department of Chemistry, University of British Columbia, 2036 Main Mall, Vancouver BC V6T 1Z1, Canada}

\correspondingauthor{Ilsa R. Cooke}
\email{icooke@chem.ubc.ca}


\begin{abstract}

Infrared emission features toward interstellar gas of the IC 348 star cluster in Perseus have been recently proposed to originate from the amino acid tryptophan. The assignment was based on laboratory infrared spectra of tryptophan pressed into pellets, a method which is known to cause large frequency shifts compared to the gas phase. We assess the validity of the assignment based on the original Spitzer data as well as new data from JWST. In addition, we report new spectra of tryptophan condensed in para-hydrogen matrices to compare with the observed spectra. The JWST MIRI data do not show evidence for tryptophan, despite deeper integration toward IC  348. In addition, we show that several of the lines attributed to tryptophan are likely due to instrumental artifacts. This, combined with the new laboratory data, allows us to conclude that there is no compelling evidence for the tryptophan assignment.

\end{abstract}

\keywords{Interstellar medium (847) --- Interstellar molecules (849) --- Astrochemistry (75) --- Laboratory astrophysics (2004) }

\section{Introduction} \label{sec:intro}


\begin{figure}
    \centering
    \includegraphics[width=0.35\textwidth]{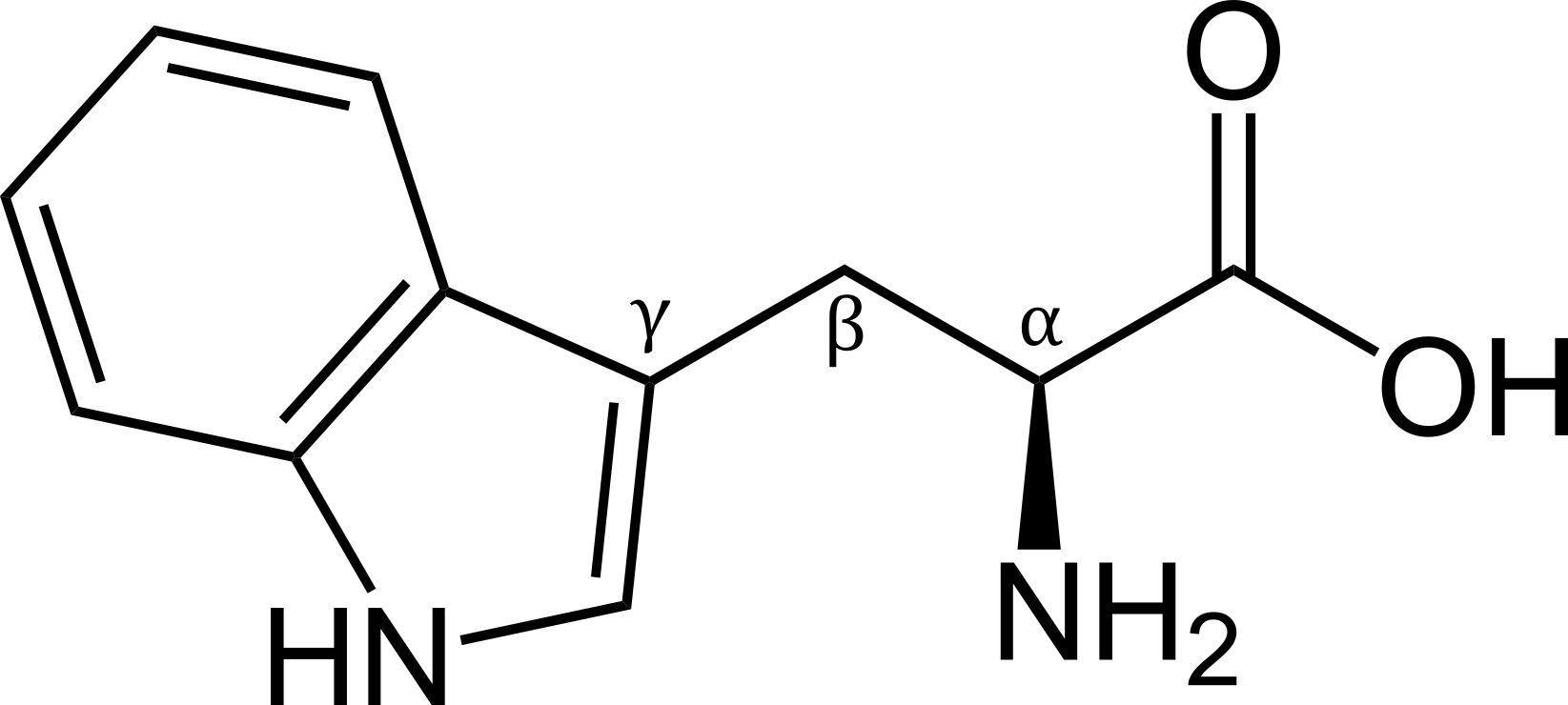}
    \caption{Structure of tryptophan (Trp). In wedge-dash notation, the dark triangle indicates that the amine (-NH$_2$) group is coming out of the page toward the reader. The carbon next to the carboxyl  (-COOH) group is called the $\alpha$-carbon, followed by the $\beta$-carbon and $\gamma$-carbon.}
    \label{fig:structure}
\end{figure}

\begin{figure*}[ht!]
    \centering
    \includegraphics[width=15 cm]{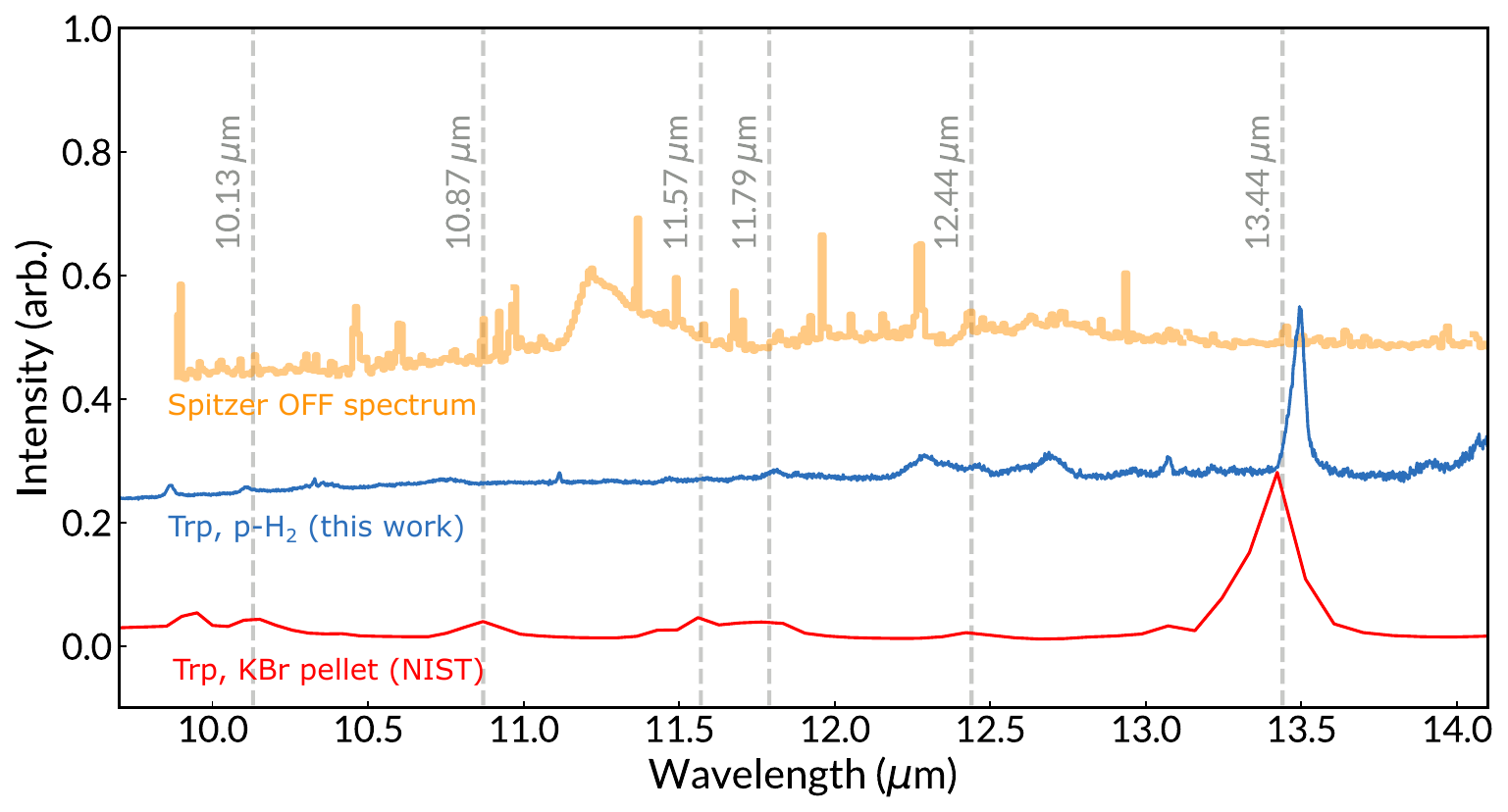}
    \includegraphics[width=15 cm]{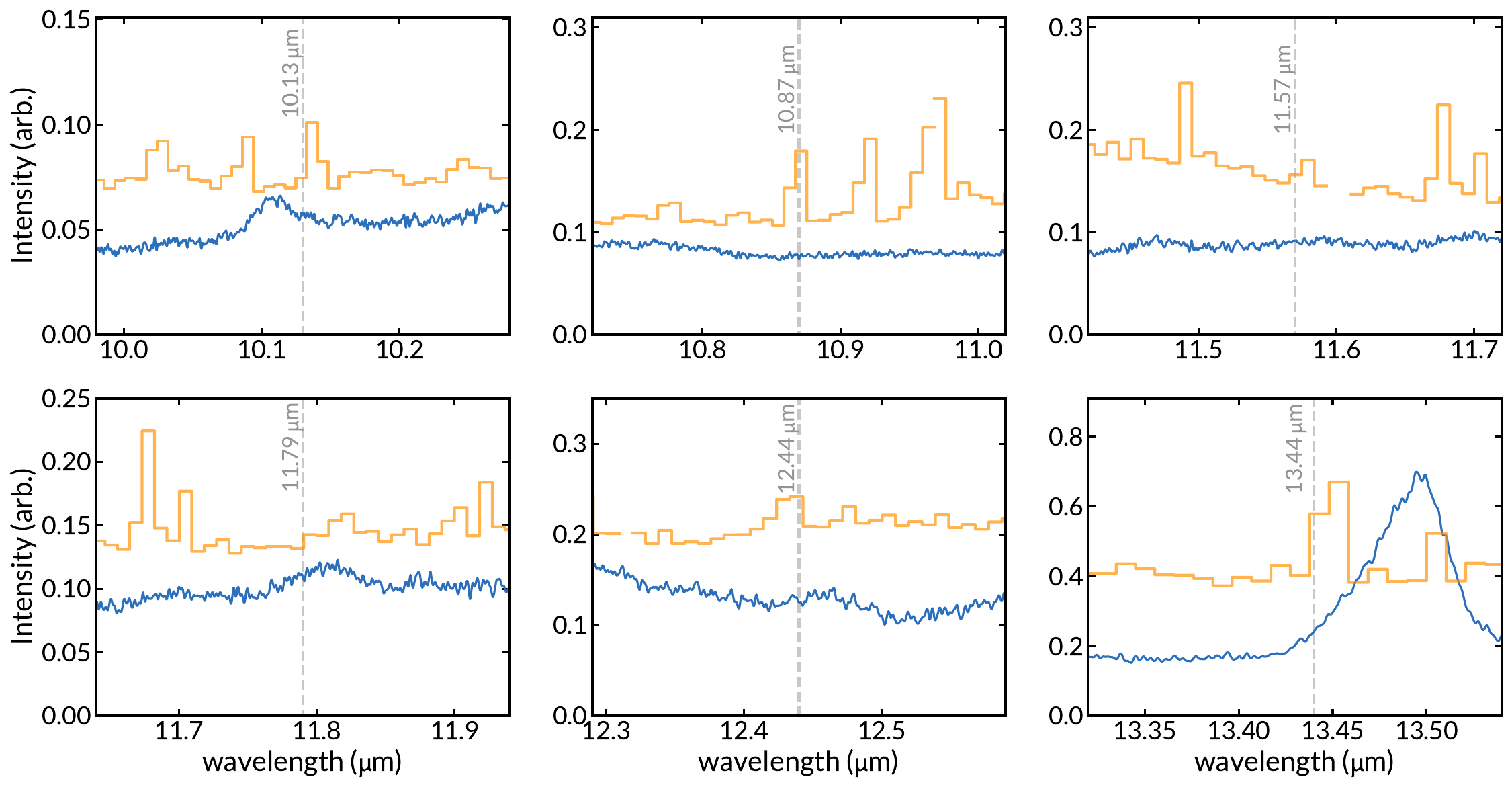}
    \caption{Comparison of the Spitzer-IRS archival spectrum of IC 348 (orange) with our laboratory spectrum of Trp isolated in p-H$_2$ (blue) and Trp in KBr pellets (red) from the NIST Chemistry Webbook. The wavelengths of the features attributed to Trp by IG23 are shown in the grey dashed lines. The Spitzer spectrum (orange) is of the dedicated background observation (AOR 22848512). Zoomed panels are centred around the features attributed to Trp by IG23.}
    \label{fig:labvsspitzer}
\end{figure*}

The search for interstellar amino acids is of importance as they are critical for life on Earth. Amino acids, molecules containing both amine (-NH$_2$) and carboxyl (-COOH) functional groups, are the building blocks of proteins, the biomolecules that provide cellular structure and function in all living organisms. Their discovery in interstellar space would have important implications for prebiotic chemistry and the origins of life. The search for the simplest amino acid glycine (\ch{NH2CH2COOH}) in the interstellar medium (ISM) started over forty years ago, soon after the detection of the first organic molecules using radio telescopes and after its rotational spectrum was characterized \citep{1978JMoSp..72..372S}; however, despite decades of searches, its detection remains elusive \citep{Brown1979, Hollis1980, Snyder1983, Combes1996, Ceccarelli2000, Synder2005, Jones2006, Cunningham2007, Jimenez-Serra20, carl2023deep,Rivilla2023}. In contrast, amino acids have been detected in meteorites, comets, and asteroids by in-situ and sample return missions  (e.g., \citealp{elsila2016meteoritic, altwegg2016prebiotic, potiszil2023insights}). 
The detection of an amino acid in the ISM would have important implications for the availability of the building blocks of life outside of our solar system.

In a recent paper, \cite{IglesiasGroth2023}, hereafter IG23, presented infrared (IR) features in archival Spitzer-IRS observations of the gas in the IC 348 star cluster in the Perseus molecular cloud and suggested that those features originate from IR emission from the complex amino acid tryptophan (\ch{C11H12N2O2}; see Figure \ref{fig:structure}); hereafter, Trp. The assignment of Spitzer data was made by comparison to the author’s laboratory spectra of Trp prepared as pressed salt pellets \citep{IglesiasGroth2021}. This method is commonly used in IR spectroscopy to analyse solid-phase samples and involves preparing pressed discs of the sample mixed with potassium bromide (KBr) or caesium iodide (CsI), since these salts are transparent in the mid-IR. 

Infrared spectroscopy provides a means to probe the vibrational modes of molecules; however, because these energy levels are spaced further apart than those of pure rotational transitions, observations of gas-phase IR emission usually require warm gas or a radiative pumping mechanism to populate the levels.
As such, the vast majority of new molecule identifications in molecular clouds are made using rotational spectroscopy in the centimeter to (sub)millimeter wavelength range, since the temperatures of these clouds can populate rotational energy levels, as opposed to vibrational levels observed in the IR \citep{mcguire2020laboratory}.

Rotational spectroscopy provides a way to uniquely and unambiguously detect and study molecules (with a non-zero electronic dipole moment) as it accurately describes a molecule's shape, size, and rotational constants.  
Some molecules, especially those not carrying a permanent dipole, have been detected in the ISM by vibrational spectroscopy, such as benzene (\ch{C6H6}) with the Infrared Space Observatory (ISO; \citealt{cernicharo2001infrared}), and recently \ch{CH3+} thanks to MIRI instrument onboard the James Webb Space Telescope (JWST; \citealt{berne2023formation}).
As such, the detection of Trp would mark a substantial advance in terms of molecular complexity, especially compared to the molecules that have been detected in the infrared \citep{McGuire2022}. In addition, while many lower mass amino acids have been detected in meteorites \citep{Burton2012}, Trp has thus far not been identified \citep{Glavin2020}. 

To reassess the claimed interstellar detection by IG23, we demonstrate that there is no evidence for Trp in data collected using the Mid-infrared Instrument (MIRI) aboard the James Webb Space Telescope (JWST). We compare the Spitzer spectrum toward IC 348 to new IR spectra of Trp isolated in solid para-hydrogen (p-H$_2$). In addition, we discuss the data reduction strategies employed by IG23 and the possibility of instrument artifacts.

Solid p-H$_2$ matrix isolation spectroscopy is an excellent technique for characterizing amino acids \citep{Momose1998MatrixIsolationSU},  providing an environment to characterize their spectroscopy and photochemistry in their neutral form, i.e., their expected form in the gas-phase of interstellar space. Compared to other methods, such as pressed pellets, or other matrices such as argon, p-\ce{H2} is advantageous due to, in general, narrower spectral linewidths and smaller frequency shifts relative to the gas phase. In addition, due to the rapid condensation of gaseous compounds in this technique, the gas-phase equilibrium conformer composition is well conserved upon deposition and is better conserved than for other matrices \citep{Wong2015}. 

Here, we report new laboratory spectra of Trp isolated in p-H$_2$ matrices and compare these spectra to the Spitzer-IRS archival data. We use data taken towards IC 348 with JWST to show that there is no evidence for Trp toward this source. Lastly, we show that several of the lines in the Spitzer data are likely due to instrumental artifacts.

\section{Experimental Methods}

The experiments were conducted using a newly constructed matrix isolation setup, similar to that described in \citet{Toh2015} and \citet{Moore2023}. Briefly, the apparatus consists of a p-H$_2$ converter, a high-vacuum chamber ($\sim$10$^{-7}$ Torr) and a high-resolution Fourier transform infrared spectrometer (FTIR).

p-H$_2$ was prepared by a similar method as mentioned in \citet{Miyamoto2012} and \citet{Tom2009}. Briefly, normal hydrogen gas of 99.99\% purity (Praxair Canada Inc.) was converted into p-H$_2$ gas, with a purity of 99.95\%, by flowing the normal gas through a magnetic catalyst (FeOH)O kept at $\sim$14 K. The p-H$_2$ was stored in stainless steel cylindrical containers before entering the vacuum chamber.

The inner matrix chamber has a ZnSe plate (Pier Optics, 2 mm thickness, wedged) fixed in the centre of a nickel-plated copper plate holder using indium, which is cooled to 3.5 K by a closed-cycle helium refrigerator (Sumitomo Heavy Industries, Ltd., SRDK-205). L-Tryptophan (Sigma Aldrich, $\geq$98\%) was directly deposited by its sublimation at 463--478 K via a Knudsen cell. The hot vapour of Trp was frozen onto the cooled window, together with a large excess of p-H$_2$.

Infrared absorption spectra were taken with a high-resolution FTIR (Bruker, IFS 120K) spectrometer equipped with a KBr beam splitter, a glowbar MIR light source, and a liquid N$_2$ cooled MCT detector. All spectra were recorded at a resolution of 0.1 cm$^{-1}$ and averaged over 50 scans.

\section{Mid-IR spectrum}

\begin{deluxetable*}{ccccc}
\tablecaption{Major bands between 1000--700 cm$^{-1}$ in the experimental spectrum of Trp isolated in a solid p-H$_2$ matrix. The differences in wavenumber (cm$^{-1}$) from those recorded by \citet{kaczor2007matrix} in solid matrices of Ar and Xe are given ($\Delta\tilde{\nu}$ = $\tilde{\nu}_{pH_2} - \tilde{\nu}_{Ar/Xe}$).
Experimental intensities (I$_{exp}$) are reported as s, m, w, or sh, where s = strong, m = medium, w = weak and sh = shoulder.\label{tab:assignments}}
\tablehead{
\colhead{$\tilde{\nu}$/cm$^{-1}$}& \colhead{$\lambda$/$\mu$m}&\colhead{I$_{exp}$}& \colhead{$\Delta\tilde{\nu}$/cm$^{-1}$ (Ar)} & \colhead{$\Delta\tilde{\nu}$/cm$^{-1}$ (Xe)}}
\startdata
       741.0  & 13.49 & s   &   1 & 2 \\
       765.1 & 13.06 & w & 0 & 0\\
       771.2 & 12.97 & w & -1 & 1 \\
    787.9   & 12.69 & m &  0 & -1\\
     802.5    & 12.46 & w &  2 & n.a.\\ 
     808.0 & 12.38 & sh &  3 & 2\\
   813.8 & 12.29 & m &  2 & 1 \\
    846.4 & 11.81 & w  & 1 & 3 \\
     855.1 & 11.69 & w & 1 & 5 \\
   862.9 & 11.59 & w &  -2 & 0 \\
    886.5 & 11.28 & w  & 2 & 2 \\
    912.2 & 10.96 & w &  -1 & 1 \\
   931.9/929.3 & 10.73/10.76 & w,w &  4/5 &  6/6 \\
   963 & 10.38 & w & 1 & 1 \\
   988.7 & 10.11 & w & 0 & 0 \\
\enddata
\end{deluxetable*}

The mid-IR spectrum from 10--14 $\mu$m of Trp isolated in solid p-H$_2$ at 3.5 K is shown in Figure \ref{fig:labvsspitzer} (blue trace). Detailed spectroscopic analysis and conformational assignment of the spectrum will be presented in a future publication. The major bands between 10--14 $\mu$m ($\sim$1000--700 cm$^{-1}$) are given in Table \ref{tab:assignments}. The strongest feature is observed at 13.49 $\mu$m ($\sim$741 cm$^{-1}$), shifted by $\sim$0.05 $\mu$m ($\sim$3 cm$^{-1}$) from that assigned to Trp by IG23. This band is also shifted to shorter wavelength by 1 and 2 cm$^{-1}$ compared to previous measurements in argon and xenon matrices, respectively \citep{kaczor2007matrix}. We observe six bands that are consistent with those of IG23 (in $\mu$m): 10.10 (10.13), 10.73/10.76 (10.87), 11.59 (11.56), 11.81 (11.79), 12.46 (12.44) and 13.49 (13.44), where the numbers in parentheses refer to those reported by IG23 based on their laboratory spectra in CsI pellets. Several other weak bands are observed, which are not reported in the laboratory spectrum of IG23 but have been previously observed in Ar/Xe matrices. Shifts on the order of 1--2 cm$^{-1}$ are observed in comparison to those reported by \citet{kaczor2007matrix} in Ar/Xe (at 11 K), except for the bands near 930 cm$^{-1}$, which are shifted by 4--6 cm$^{-1}$. Larger shifts are observed in comparison to the spectrum of Trp in a KBr pellet, obtained from NIST (Figure \ref{fig:labvsspitzer}, red trace), which were likely obtained at room temperature. The width of the bands is influenced by the population of different conformers as well as inhomogeneities in the local matrix environment. The weak interaction between Trp and the p-H$_2$ host reduces the effects of trapping site inhomogeneities, resulting in the observed sharper absorption linewidths in comparison to salt pellets.

\citet{Cao1999} measured the IR spectrum of Trp isolated in KBr by spraying an aqueous solution of Trp and KBr onto an IR transparent window at 78 $^\circ$C. In this way, the sample molecules are isolated from one another in the KBr matrix. The authors observe a strong mode at $\sim$13.5 $\mu$m ($\sim$740 cm$^{-1}$). 

\textit{Ab initio} calculations show that the observed bands are due to a convolution of several different conformers of Trp, which are present due to the flash condensation of sublimated Trp onto the cold window used for matrix isolation spectroscopy. 

Previous \textit{ab-initio} studies of Trp's conformer populations have been carried out at a variety of levels of theory \citep{Narahari14,Kim11,Jarrold01,Lin2005,Fausto07,Simons01}, including a recent study that implements high-level coupled cluster (CCSD(T)/aug-cc-pVDZ) single point energy corrections \citep{Chen21}. 

A prior high-level analysis of the relative stability of the conformers of Trp was carried out in \citet{Bouchoux12}.
From the highest-level calculations (G4(MP2), 18 of the 20 lowest energy conformers have $>$1\% population at 298 K with the lowest energy conformer accounting for 30\% and 9 conformers, each having between 2 and 5\% population, contributing a total of $\sim$30\%. At 475 K, the relative population of the less populated conformers compared to the lowest energy conformer will increase further. 
While these studies present differences in the order of the relative energies of the conformers, all of these works highlight that at elevated temperatures, a large number of conformers will be populated. 
Spectra of gas-phase Trp recorded by \citet{Bakker2003} using infrared ion-dip spectroscopy showed that the conformation has a large effect on the vibrational structure. Further laboratory work is necessary before Trp can be searched for using infrared telescopes.

\section{JWST observation of IC 348}

\begin{figure*}
    \centering
    \includegraphics[width=1.0\textwidth]{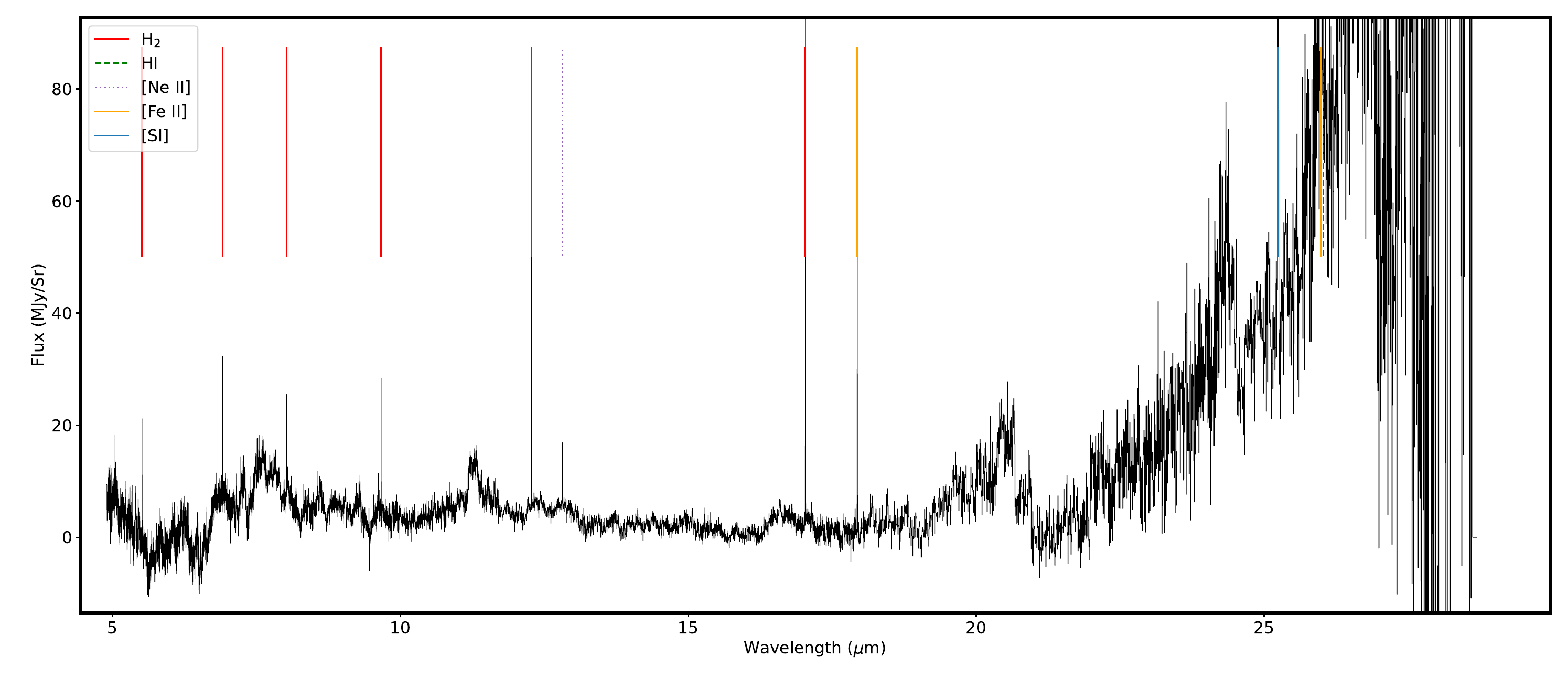}
    \includegraphics[width=1.0\textwidth]{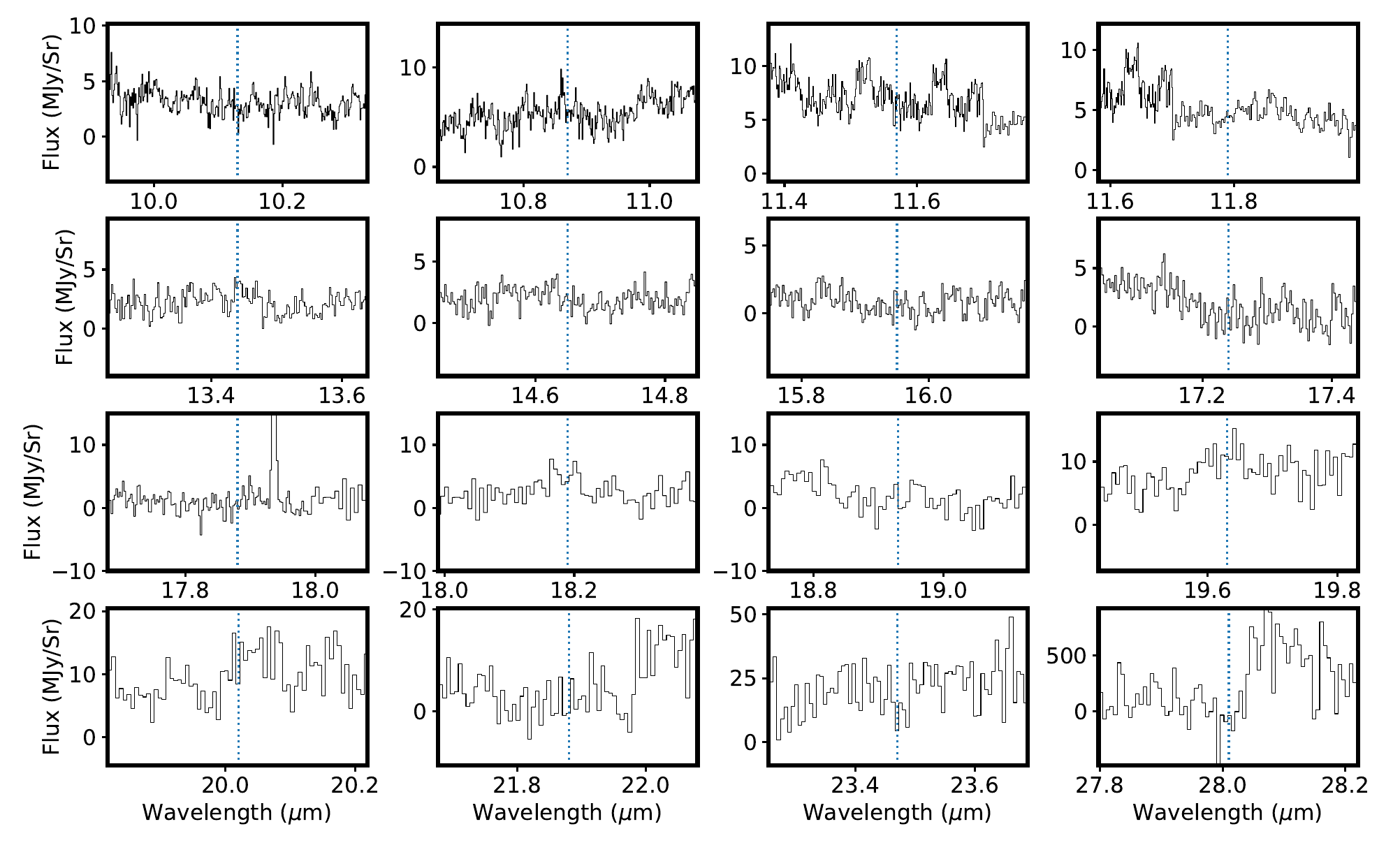}
    \caption{Top: JWST-MIRI IFU spectrum of IC348. Bottom: Zoom-in panels of the MIRI Spectrum of IC348 at the wavelengths of the lines reported by IG23 and attributed to Tryptophan.Nearby transitions corresponding to atomic and molecular H$_2$ lines are indicated by vertical coloured bars.}
    \label{fig:miri-spec}
\end{figure*}

IC 348 was observed with JWST-MIRI integral field units (IFUs) as part of the Guaranteed time observations (GTO) program PID 1236 (PI Michael Ressler). One of the pointings in this program is centered on equatorial coordinates, J2000 (3:43:56.980, +32:03:03.98), which is close to (within a few arcminutes of) the positions observed with Spitzer in \citet{IglesiasGroth2023}, that is RA: (03:43:50.10, 03:45:00.87), Dec: (+31:59:54.1, +32:09:15.5). We retrieved the level 3 MIRI-IFU cube for this pointing from the MAST online archive.\footnote{https://mast.stsci.edu}
In the IFU cube, no point source is present, indicating that the observed emission is largely dominated by the Perseus cloud gas, where Trp was claimed to be detected by IG23.
We extract a spectrum from the IFU cube using a circular aperture of radius 1.15'', centered on coordinates (3:43:56.980,+32:03:03.98).
The resulting spectrum is shown in Fig.~\ref{fig:miri-spec}.
Polycyclic aromatic hydrocarbon (PAH) emission is observed at 7.7 $\mu$m and 11.2 $\mu$m as reported by IG23. 
Several H$_2$ pure rotation lines, HI recombination lines, and atomic fine-structure lines are also present, confirming that nebular gas is well detected.  
We have searched for the lines claimed to be detected by IG23 using these data. In the bottom panels of Fig.~\ref{fig:miri-spec}, we show extracts of the spectrum, centered on the wavelengths of the lines attributed to Trp by IG23. None of the lines are detected in the MIRI data. We compute the detection limits at the expected position, using the local root mean square noise and a line width 0.003 $\mu$m, which is the measured FWHM of the HI 9-13 line at 9.392 $\mu$m.
In Table \ref{tab:JWST}, we compare those detection limits to the intensities reported by IG23. 
For all lines below 28 $\mu$m, the JWST detection limits are well below the intensities reported by IG23. Above 28 $\mu$m, the JWST detection limit is below that of Spitzer, first because the aperture of MIRI is much smaller (about 50 arcsec$^2$ for MIRI-IFU, versus 
nearly 250 arcsec$^2$ for Spitzer-IRS) and the intrinsic sensitivity of MIRI decreases drastically at long wavelengths.\footnote{https://jwst-docs.stsci.edu/jwst-mid-infrared-instrument/miri-performance/miri-sensitivity} 
However, we can firmly reject the detection of the lines below 28$\mu$m, where the limit of detection is a up to factor ten times smaller than the reported Spitzer intensities of IG23.

\section{Origin of the lines observed by Spitzer}

The identification of Trp presented in \citet{IglesiasGroth2023} relies on a series of IR lines in Spitzer-IRS spectra of IC-348 cloud in Perseus. To identify potential sources of contamination in the data leading to a false detection of lines, we obtained a sample of the Spitzer data used by IG23.
More specifically, we retrieved from the Spitzer Heritage Archive the basic calibrated data (BCD) for the Astronomical Observation Request (AOR) 22848512 used by IG23. 
This AOR was originally designed to constitute a background observation, to be subtracted from a science observation of a young stellar object with AOR 22848256, which we also retrieved (not used in IG23). These data are taken from observing program 40247 (PI. N. Calvet) entitled ``Probing the Gas in the Planet Forming Regions of Protoplanetary Disk.''
The spectra from these two positions (separated by about 10'') are presented in Figure~\ref{fig:spitzer_pixels}, where the science observation is the blue trace and the background position is the orange trace. 
Here, we used the CUBISM software \citet{smith2007spectral} to reconstruct the cubes from the BCD files the spectra are then extracted over the full aperture of the IRS spectrograph slits.

In both spectra, dust continuum emission and broad bands (at 11.2 and 12.7 $\mu$m notably) associated with the emission of polycyclic aromatic hydrocarbons (PAHs) are seen. In addition, numerous lines are observed, some of which (labeled on Figure ~\ref{fig:labvsspitzer} and Figure~\ref{fig:spitzer_pixels}) were attributed to Trp by IG23. 

The common practice when deriving Spitzer spectra is to subtract the background observation from the science observation.
This allows the removal of rogue pixels in the Spitzer IRS detectors. A rogue pixel is a pixel with abnormally high dark current and/or photon responsivity that manifests as pattern noise in a Spitzer IRS basic calibrated data image. 

In Figure ~\ref{fig:spitzer_pixels}, we show the spectrum resulting from the science minus background subtraction (green trace). When performing this subtraction, numerous lines are removed, notably most of the lines attributed to Trp. This indicates a systematic ``noise'' due to rogue pixels creating artificial lines. Inspection of the individual detector frames in the level 1 Spitzer-IRS data for these observations provides additional evidence of rogue pixels (see Appendix, Figure \ref{fig:frames}).
Overall, there is strong evidence that several of the lines attributed to Trp by IG23 could be due to instrumental artifacts in Spitzer observations. IG23 relied on spectra available in the Cornell Atlas of Spitzer/Infrared Spectrograph sources (CASSIS). These spectra are derived using an automatic procedure which does not include background subtraction. The data processing in CASSIS does include an automatic filtering procedure allowing removal of the strongest rogue pixels, however, it is not suited to remove low-level rogue pixels \citep{lebouteiller2015cassis,lebouteiller2023}.

\section{Discussion and Astrophysical Implications}

The confirmed detection in the gas phase of an interstellar amino acid would have substantial implications from both an astrochemical and astrobiological standpoint. While IR spectroscopy has characterized some astronomical features from large molecules such as PAHs, the only confirmed detections of individual PAH molecules in astronomical environments have been by rotational spectroscopy at radio/mm-wavelengths \citep{McGuire2021PAHs,Burkhardt2021, Cernicharo2021, Sita_2022}.  As such, the identification of Trp would represent one of the biggest jumps in molecular complexity detected in astronomical environments. To date, neither of the backbones of Trp --- indole (C$_8$H$_7$N) and alanine (CH$_3$CHNH$_2$COOH) --- have been detected. 

Recently, \citet{Hudson2023} re-examined the assignment of Trp made by IG23. They discussed several problems with the assignment, including the use of the solid-phase absorption wavelengths to assign gas-phase emission. In addition, they discuss the differences in molecular structure that are expected to be present in the solid versus the gas phase. Trp in the gas phase is found in its neutral form, while in solid films, Trp adopts the zwitterionic form. Importantly, amino acids isolated in p-H$_2$ are typically in their neutral form, thus permitting better comparison to the gas phase. Lastly, \citet{Hudson2023} discussed concerns regarding the derivation of the band strengths used to calculate the molecular abundance. Our findings, alongside those of \citet{Hudson2023}, demonstrate that the IR features in the Spitzer spectra of IC 348 cannot be assigned to Trp. Future IR searches, for example with JWST, would require high-resolution gas phase mid-IR spectra, which are difficult to obtain and assign. 

\begin{figure*}
    \centering
    \includegraphics[width=14cm]{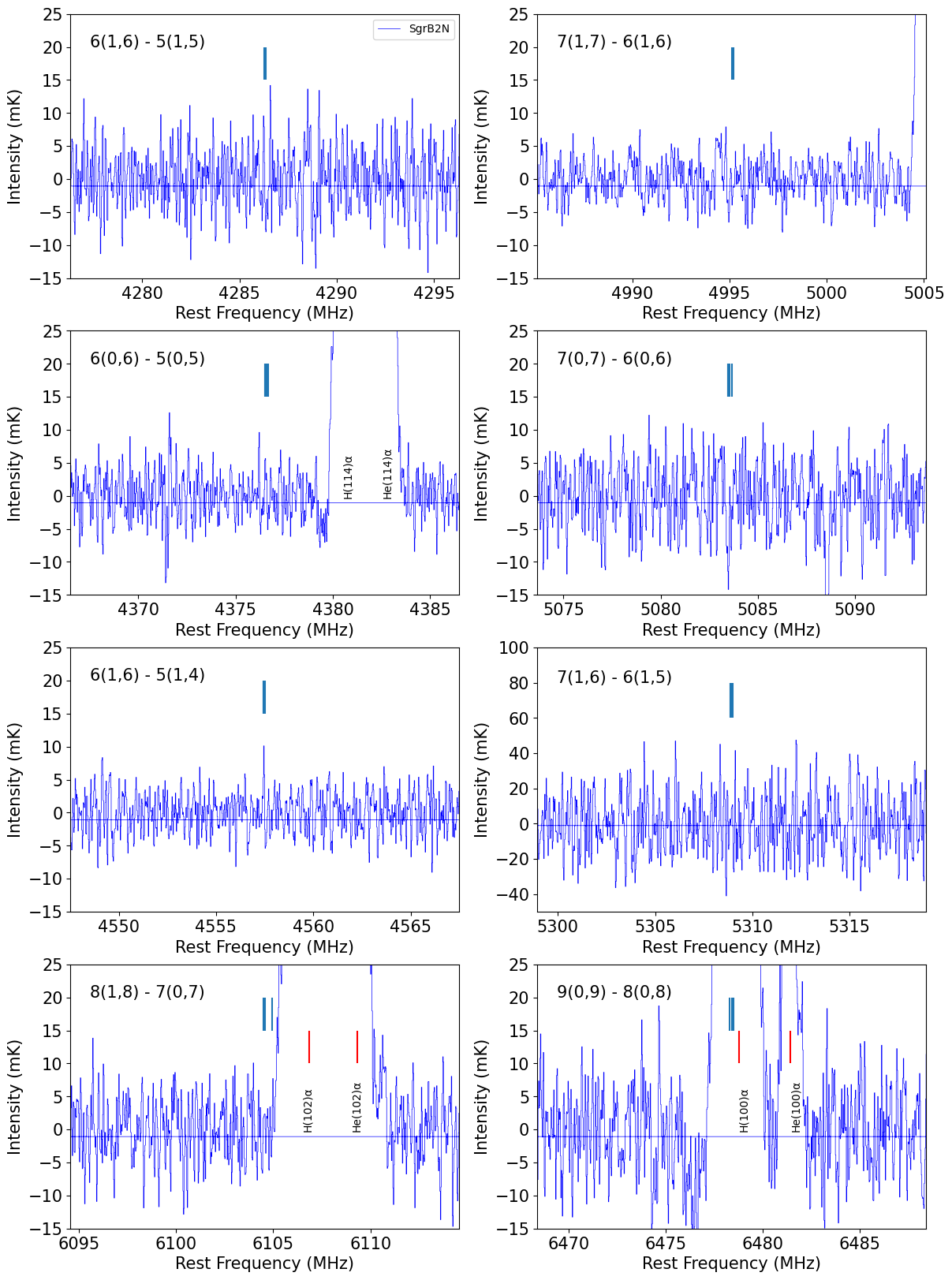}
    \caption{Observed spectra of the PRIMOS observations (blue trace) toward Sgr B2(N) near the reported rest frequencies of Trp.  Vertical marks near the center of each panel correspond to the Trp laboratory rest frequencies report in Table S2 of \citet{Sanz2014}.  Transition quantum numbers (without hyperfine splitting and quadrupole coupling) are given in the top left of each panel.  Nearby transitions belonging to radio recombination lines (red vertical marks) are labeled.}
    \label{fig:GBT_spectra}
\end{figure*}

Laboratory rotational spectra of some low-energy conformers of Trp have been reported permitting future searches at mm/sub-mm wavelengths. \citet{Sanz2014} report a gas-phase rotational spectrum of Trp, where several low-lying transitions were measured in the laboratory. While a detailed search and upper limit analysis is outside the scope of this work, we conducted a cursory search in the archival astronomical database the PRebiotic Interstellar MOlecule Survey (PRIMOS)\footnote{\label{fn:primos}Access to the entire PRIMOS data set, specifics on the observing strategy, and overall frequency coverage information is available at \url{http://archive.nrao.edu} by searching for GBT Program ID: AGBT07A\_051} on the GBT.  Figure \ref{fig:GBT_spectra} illustrates the overlap in coverage between the laboratory data and the GBT astronomical spectrum.  The vertical fiducial marks near the center of each spectral window indicate the rest frequencies of the spectral lines reported in Table S2 of \citet{Sanz2014}.  The transition quantum numbers are reported at the top left of each panel\footnote{\citet{Sanz2014} do report additional quantum numbers to account for hyperfine splitting and quadrupole coupling that are represented by the vertical marks but not labeled as part of the quantum numbers at the top left of the panel}.  The 1$\sigma$ rms noise level in each spectral window is $\sim$3 mK except for the 7(1,6) - 6(1,5) passband where the noise level was higher ($\sim$12 mK).  No spectral features were detected at the rest frequencies of Trp reported by \citet{Sanz2014}. We note that the signal observed coincident with the frequency of the 6(1,6)--5(1,4) transition is not a spectral line, since it does not span multiple channels.

Lastly, amino acids may be present in the condensed phase, on icy dust grains in the ISM. Infrared spectra of low-temperature amino acid films have been measured previously, which may guide future searches and laboratory work \citep{Gomez-Zavaglia2003}. \citet{Mate2011}  reported solid-phase spectra of glycine mixed with major ice components (e.g., H$_2$O, CO$_2$) but to our knowledge, no spectra of Trp have been reported for realistic interstellar ice analogues.

\section{Conclusions}

We report new spectroscopic measurements of Trp in solid p-H$_2$ matrices. Combined with JWST data towards IC 348, we do not find any evidence for Trp in the spectrum of IC348 and conclude that the assignment of the IR emission to Trp is not supported. Follow-up laboratory measurements of the vibrational and rotational spectrum of gas-phase tryptophan are required to facilitate future searches.

\section{Acknowledgements}

We thank Reggie Hudson for useful discussions. I.R.C. acknowledges support from the University of British Columbia, the Natural Sciences and Engineering Research Council of Canada, the Canada Foundation for Innovation and the B.C. Knowledge Development Fund (BCKDF). The National Radio Astronomy Observatory is a facility of the National Science Foundation operated under cooperative agreement by Associated Universities, Inc.  The Green Bank Observatory is a facility of the National Science Foundation operated under cooperative agreement by Associated Universities, Inc.
V.M.R. acknowledges support from the grant number RYC2020-029387-I funded by MICIU/AEI/10.13039/501100011033 and by "ESF, Investing in your future"; from the grant No. PID2022-136814NB-I00 by the Spanish Ministry of Science, Innovation and Universities/State Agency of Research MICIU/AEI/10.13039/501100011033 and by ERDF, UE; and from the Consejo Superior de Investigaciones Cient{\'i}ficas (CSIC) and the Centro de Astrobiolog{\'i}a (CAB) through the project 20225AT015 (Proyectos intramurales especiales del CSIC).

\bibliography{refs}{}
\bibliographystyle{aasjournal}





\appendix

\section{Spitzer Data}

\begin{figure*}[h]
    \centering
    \includegraphics[width=15 cm]{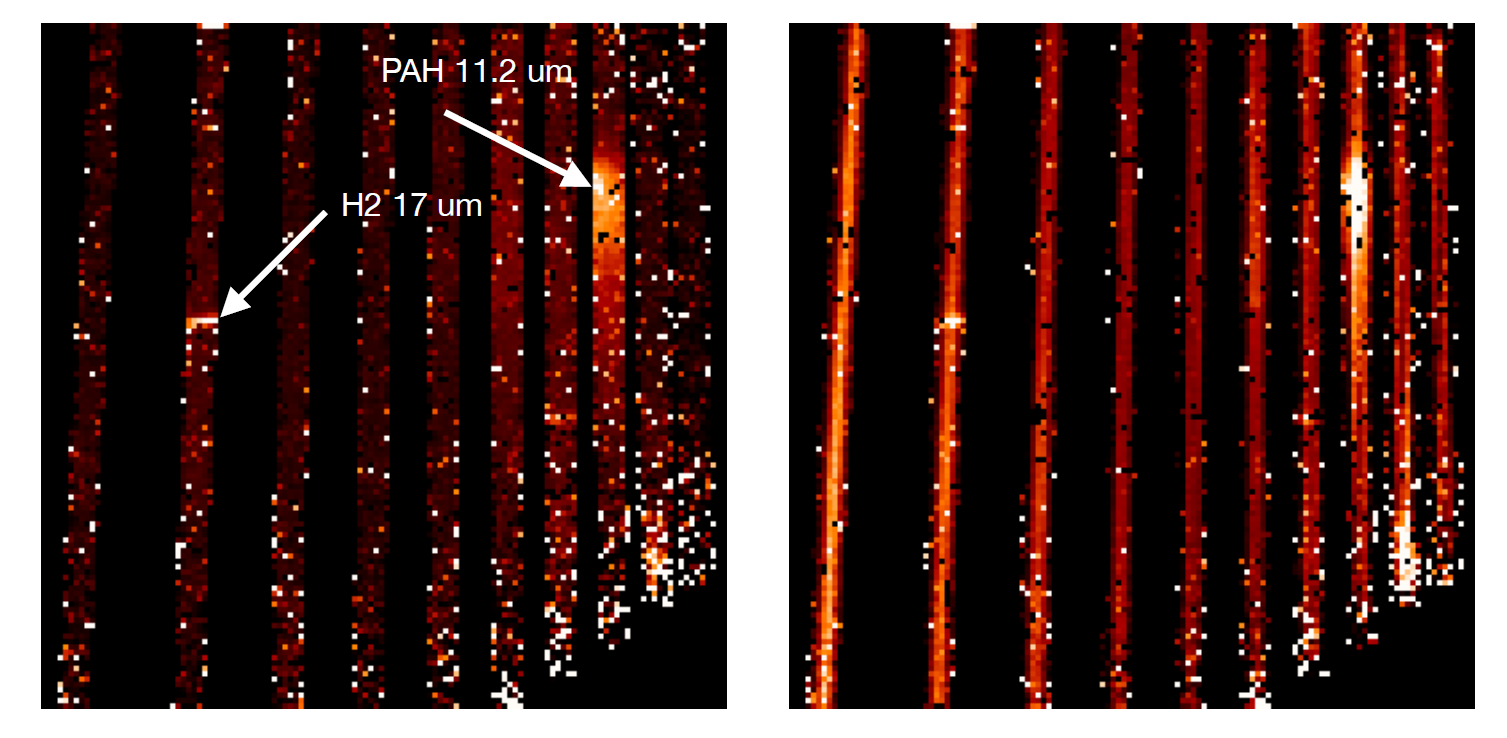}
    \caption{Averaged post BCD frames of the IRS SH observations of IC 438 at off position (left) used by the authors, and on position (right). Spectral features of H$_2$ (17 $\mu$m) and PAHs (11.2 $\mu$m) are clearly seen. Several bright ``rogue pixels'' of the detector are present in both observations, demonstrating the need to subtract the OFF from the ON position to remove the emission from these pixels.}
    \label{fig:frames}
\end{figure*}

The likelihood that some of the lines identified by IG23 are in reality due to rogue pixels of the detector is further supported by looking at the basic calibrated data (BCD) frames in the Spitzer archive. Figure \ref{fig:frames} shows an average of 30 integrations from AOR 22848512
(off position) and 60 integrations from AOR 22848256 (on position). Many bright ``rogue'' pixels in the detector are seen, and present in both the ON and OFF positions. If the OFF is not subtracted from the ON, these pixels will contribute by creating artificial lines in the spectra, unless a rigorous procedure of cleaning is performed.

\begin{figure*}[ht!]
    \centering
    \includegraphics[width=16cm]{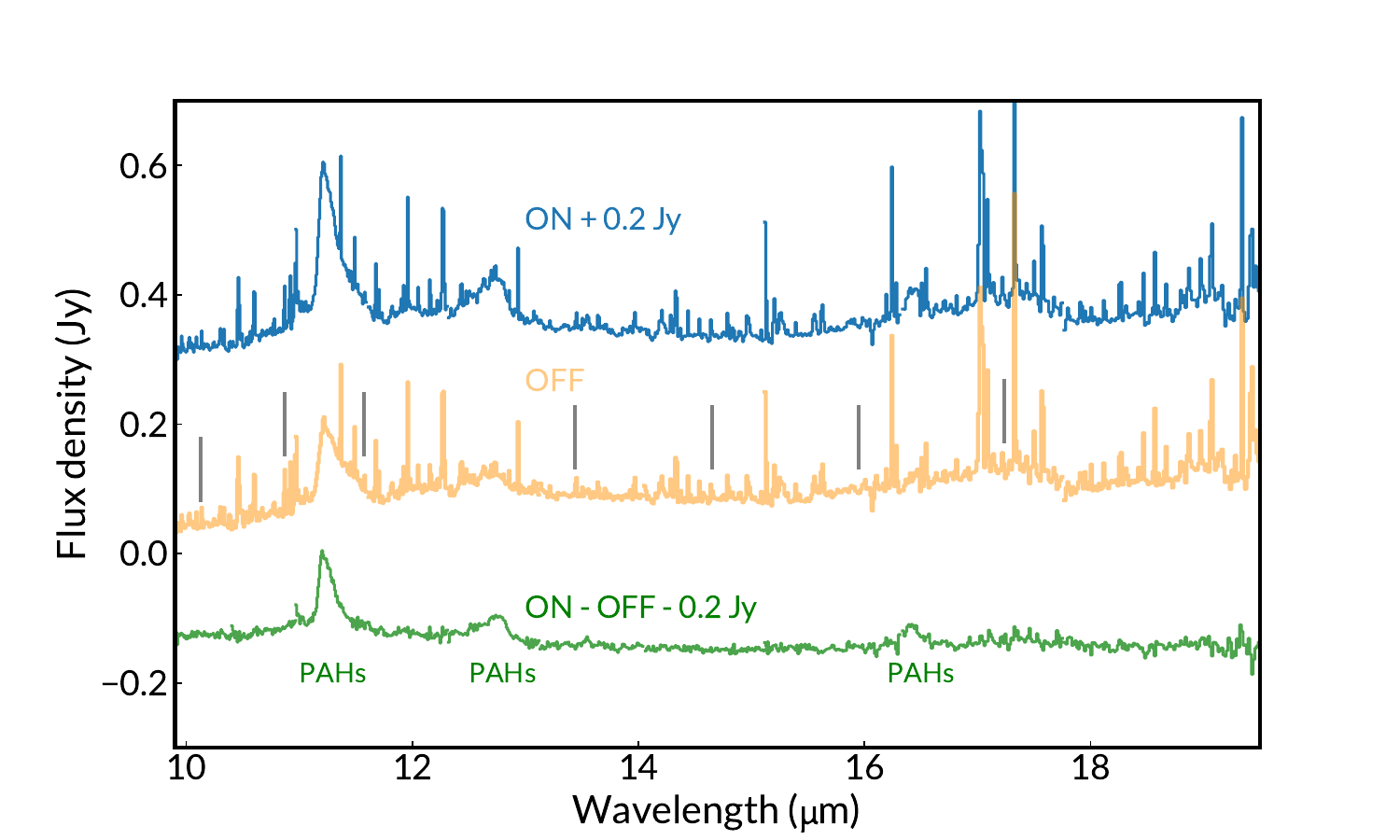}
    \includegraphics[width=16cm]
    {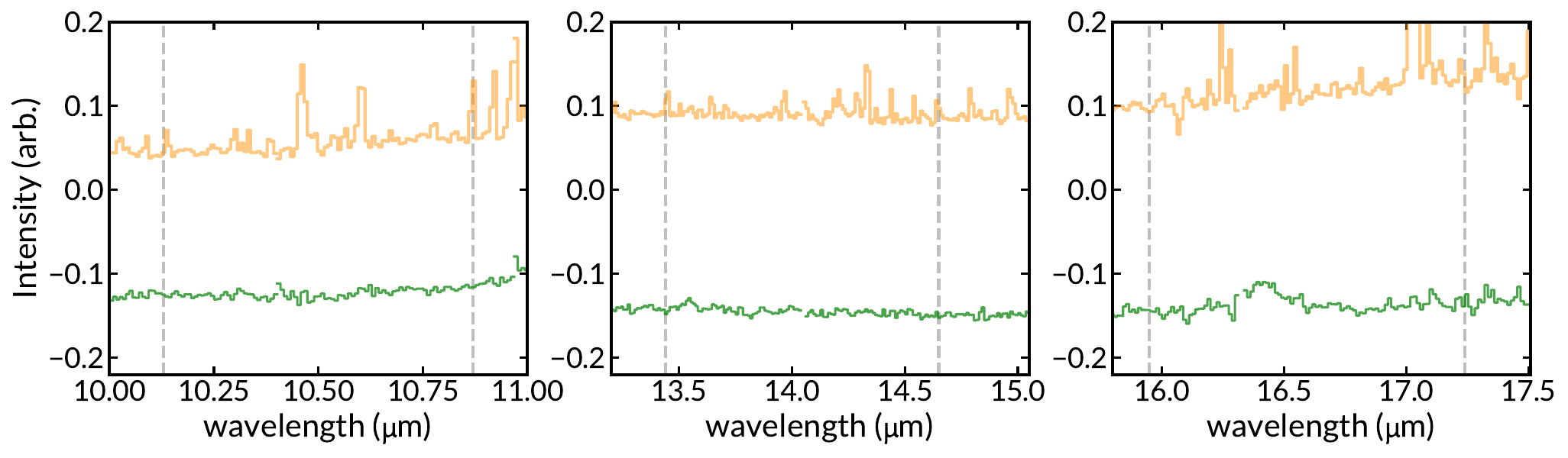}
    \caption{Spitzer-IRS medium resolution spectra obtained from archival data. The blue line shows the spectrum extracted on the science target (AOR 22848256) offset by 0.2 Jy for clarity. The orange line is the spectrum of the dedicated background observation (AOR 22848512). Numerous sharp lines are seen. The strongest lines attributed to the amino acid Trp by IG23 in the OFF spectrum are indicated with vertical grey lines. The green spectrum shows the result of ON - OFF (offset by 0.2 Jy), showing that many lines disappear including those attributed to Trp. The panels below show expansions of the data centred around the wavelengths of the signals attributed to Trp by IG23 (grey dashed lines). Broad features observed at $\sim$11.2, 12.7 and 16.4 $\mu$m are due to PAHs.}
    \label{fig:spitzer_pixels}
\end{figure*}

\section{JWST Data}

To compute the intensity from IG23, we use their fluxes and an aperture of 55.11 arcseconds corresponding to IR-SH for lines below 19 $\mu$m, and an aperture of 247.53 arcsec$^{2}$ beyond, corresponding to the IRS-LH. Comparisons between their intensities and the MIRI detection limit are given in Table \ref{tab:JWST}.

\begin{deluxetable*}{ccc}
\tablecaption{Comparison of the reported intensities from IG23 to the JWST-MIRI detection limits.\label{tab:JWST}}
\tablehead{
\colhead{Wavelength/$\mu$m} & \colhead{Intensity IG23 (10$^{-10}$ W/m$^2$/Sr)} &\colhead{MIRI Detection limit (10$^{-10}$ W/m$^2$/Sr)}}
\startdata
       10.13  & 11 & 0.87  \\
       10.87 &  15 & 0.98\\
       11.57 &  24 & 1.0 \\
       11.79 &  11 & 0.74\\
       13.44 &  35  & 0.50 \\
       14.65 &   7.6 & 0.36\\
       15.95 &   8.1 & 0.28\\
    17.24 &   24 &  0.44 \\
  17.88 &   18&   2.0\\
    18.19 &   11 &  0.51\\
    18.93 &   6.5 &  0.47 \\
  19.63  & 3.6& 0.63 \\
20.02 &   1.8 &  0.88 \\
21.88 &   2.4 &  0.68 \\
28.01 & 1.8 & 44.7\\  
28.70 & 3.6 & 28.8 \\
\enddata
\end{deluxetable*}

\end{document}